\documentclass[a4paper,10pt]{article}

\usepackage{epsf}
\usepackage{graphicx}    

\newcommand{\be}[1]{\begin{equation}\label{#1}}
\newcommand{\ee}{\end{equation}}
\newcommand{\bea}{\begin{eqnarray}}
\newcommand{\eea}{\end{eqnarray}}

\renewcommand{\markright}{\markright{\thepage}}

\begin{document}

\begin{titlepage}

\begin{flushright}
astro-ph/0512018
\end{flushright}

\vspace{5mm}

\begin{center}

{\Large \bf A Note on Crossing the Phantom Divide in Hybrid Dark
Energy Model}

\vspace{10mm}

{\large Hao Wei\footnote{\,email address: haowei@itp.ac.cn}}

\vspace{5mm} {\em  Institute of Theoretical Physics, Chinese
Academy of Sciences,\\
 P.O. Box 2735, Beijing 100080, China \\
 Graduate School of the Chinese Academy of Sciences, Beijing 100039, China}

\vspace{10mm}
 {\large Rong-Gen Cai\footnote{\,email address: cairg@itp.ac.cn}}

\vspace{5mm}
 {\em  Institute of Theoretical Physics, Chinese
Academy of Sciences,\\
 P.O. Box 2735, Beijing 100080, China}

\end{center}

\vspace{5mm}
\begin{abstract}
Recently a lot of attention has been given to building dark energy 
models in which the equation-of-state parameter $w$ can cross the 
phantom divide $w=-1$. However, to our knowledge, these models 
with crossing the phantom divide only provide the possibility that 
$w$ can cross $-1$. They do not answer another question: {\em why 
crossing phantom divide occurs recently?} Since in many 
existing models whose equation-of-state parameter can cross the 
phantom divide, $w$ undulates around $-1$ randomly,  {\em why 
are we living in an epoch $w<-1$?} This can be regarded as the 
second cosmological coincidence problem. In this note, we propose 
a possible approach to alleviate this problem within a hybrid dark 
energy model.\\

\end{abstract}

\end{titlepage}

\newpage

\setcounter{page}{2}


Many cosmological observations, such as SNe Ia~\cite{r1}, WMAP~\cite{r2}, 
SDSS~\cite{r3}, Chandra X-ray Observatory~\cite{r4} etc., discover that our 
universe is undergoing an accelerated expansion. They also suggest that our 
universe is spatially flat, and consists of about $70\%$ dark energy with 
negative pressure, $30\%$ dust matter (cold dark matters plus baryons), 
and negligible radiation. Dark energy has been one of the most active fields 
in modern cosmology~\cite{r5}. 

In the observational cosmology of dark energy, the 
equation-of-state parameter (EoS) $w\equiv p/\rho$ plays a central 
role, where $p$ and $\rho$ are its pressure and energy density, 
respectively. To accelerate the expansion, the EoS of dark energy 
must satisfy $w<-1/3$. The simplest candidate of the dark energy 
is a tiny positive time-independent cosmological constant 
$\Lambda$, for which $w=-1$. However, it is difficult to 
understand why the cosmological constant is about 120 orders of 
magnitude smaller than its natural expectation, i.e. the Planck 
energy density. This is the so-called cosmological constant 
problem. Another puzzle of the dark energy is the cosmological 
coincidence problem: {\em why are we living in an epoch in which 
the dark energy density and the dust matter energy density are 
comparable?} This problem becomes very serious especially for the 
cosmological constant as the dark energy candidate. The 
cosmological constant remains unchanged while the energy densities 
of dust matter and radiation decrease rapidly with the expansion 
of our universe. Thus, it is necessary to make some fine-tunings. 
In order to give a reasonable interpretation to the cosmological 
coincidence problem, many dynamical dark energy models have been 
proposed as alternatives to the cosmological constant. The famous 
one is quintessence~\cite{r6,r7,r36}, a cosmic real scalar field that 
is displaced from the minimum of its potential. With the evolution 
of the universe, the scalar field slowly rolls down its potential. 
The Lagrangian density for the quintessence is 
\be{eq1}
{\cal L}_{quintessence}=\frac{1}{2}\left(\partial_\mu
\varphi\right)^2-V(\varphi),
\ee
where $\varphi$ is a real scalar field, and we adopt the metric 
convention as $(+,-,-,-)$ and use the units $\hbar=c=8\pi G=1$ 
throughout this note. A class of tracker solutions of quintessence 
is found in order to solve the cosmological coincidence 
problem~\cite{r7}. Another famous dark energy candidate is 
phantom~\cite{r8,r9,r10,r37}, which has a ``wrong'' sign kinetic 
energy term, i.e. 
\be{eq2}
{\cal L}_{phantom}=-\frac{1}{2}\left(\partial_\mu
\varphi\right)^2-V(\varphi). 
\ee 
It is easy to see that the EoS of quintessence (phantom) is always 
larger than (less than) $-1$.

Recently, by fitting the SNe Ia data, marginal evidence for the 
EoS of dark energy $w(z)<-1$ at redshift $z<0.2$ has been 
found~\cite{r11}. In addition, many best-fits of the present value 
of $w$ are smaller than $-1$ in various data fittings with 
different parameterizations (see~\cite{r12} for a recent review). 
The present data seem to slightly favor an evolving dark energy 
with $w$ crossing $-1$  from above to below in the near 
past~\cite{r13}. Obviously, the EoS $w$ cannot cross the so-called 
phantom divide $w=-1$ for quintessence or phantom alone. 
Some efforts~\cite{r14,r15,r16,r17,r18,r19,r20,r21,r22,r23,r24,r25,r26,r27,
r28,r29,r34,r35} have been made to build dark energy model 
whose EoS can cross the phantom divide. However, to our knowledge,
 many of the existing models only provide the 
possibility that $w$ can cross $-1$. They do not answer another 
question: {\em why crossing phantom divide occurs recently?} Since 
in many existing models whose EoS can cross the phantom divide, 
$w$ undulates around $-1$ 
randomly, {\em why are we living in an epoch $w<-1$?} It can 
be regarded as the second cosmological coincidence problem.

In this note, we propose a possible approach to alleviate the 
so-called second cosmological coincidence problem. The key point 
is the {\em trigger} mechanism. It is reminiscent of the 
well-known hybrid inflation~\cite{r30}, which arises naturally in 
many string/brane inspired inflation models (see e.g.~\cite{r31} 
for a comprehensive review). In the hybrid inflation model, the 
effective potential is given by 
\be{eq3}
V(\phi,\sigma)=\frac{1}{2}m^2\phi^2+\frac{g^2}{2}\phi^2\sigma^2+
\frac{1}{4\lambda}\left(M^2-\lambda\sigma^2\right)^2, 
\ee 
where $\phi$ and $\sigma$ are two canonical real scalar fields, and 
play the roles of inflaton and the ``waterfall'' field, respectively. 
The feature of spontaneous symmetry breaking in the hybrid 
inflation model is critical. At the first stage, the field 
$\sigma$ is trapped to $\sigma=0$, while the field $\phi$ slowly 
rolls down and drives the inflation. When the field $\phi$ reaches 
a critical value $\phi_c$, the field $\sigma$ is triggered and 
starts to roll down rapidly. And then, the inflation ends. 
Recently, Gong and Kim in~\cite{r32} have proposed a model, which 
can describe simultaneously the primordial inflation and present 
accelerated expansion of the universe (see~\cite{r38} also), 
by employing a mechanism 
similar to the hybrid inflation. In their model, three canonical 
real scalar fields $\phi$, $\psi$ and $\sigma$ appear, and play 
the roles of inflaton, transition field and quintessence, 
respectively. The effective potential is given by 
\be{eq4}
V(\phi,\psi,\sigma)=\frac{1}{2}m^2\phi^2+\frac{g^2}{2}\phi^2\psi^2
-\frac{h^2}{2}\psi^2\sigma^2+\frac{1}{4\lambda}\left(M_{\psi}^2-
\lambda\psi^2\right)^2+\frac{1}{4\mu}\left(M_{\sigma}^2+
\mu\sigma^2\right)^2.
\ee
 It is worth noting that the third and 
fifth terms involved $\sigma$ are different from the other terms 
involved $\phi$ and $\psi$, which are similar to those in the 
hybrid inflation model. As was argued in~\cite{r32}, this kind of 
effective potential may arise from the supersymmetric theories. In 
this model, there are three different stages. At the first stage, 
the fields $\sigma$ and $\psi$ are trapped to $\sigma=0$ and 
$\psi=0$, while the field $\phi$ slowly rolls down and drives the 
primordial inflation. When $\phi$ reaches a critical value 
$\phi_c$, the transition field $\psi$ is triggered to roll down 
and the primordial inflation ends. When $\psi$ rolls down to a 
critical value $\psi_c$, the quintessence field $\sigma$ is 
triggered to slowly roll down its potential and drives the present 
accelerated expansion.

Enlightened by the hybrid inflation model~\cite{r30} and the model 
by Gong and Kim~\cite{r32}, we find that the trigger mechanism can 
be employed to alleviate the so-called second cosmological 
coincidence problem mentioned above. Since our interest is the 
issue of dark energy problem,  here we consider a model with two 
real scalar fields, rather than three real scalar fields 
like~\cite{r32}. Note that the fields appear in either the hybrid 
inflation model~\cite{r30} or the work of Gong and Kim~\cite{r32} 
are all quintessence-like canonical real scalar fields, namely 
their kinetic energy terms are all positive. Thus, their EoS are 
always larger than $-1$. However, our main aim here is to 
implement a mechanism for crossing the phantom divide. To this 
end, we employ a quintessence and a phantom in our model. The 
effective potential in our model is
\be{eq5}
V(\phi,\sigma)=\frac{1}{2}m^2\phi^2-\frac{g^2}{2}\phi^2\sigma^2+
\frac{1}{4\lambda}\left(M^2+\lambda\sigma^2\right)^2,
\ee
where $m$, $M$, $g$ and $\lambda$ are all positive constants; 
the field $\phi$ is a quintessence with kinetic energy term 
$\dot{\phi}^2/2$, the field $\sigma$ is a phantom with kinetic 
energy term $-\dot{\sigma}^2/2$, and we have supposed that they 
are spatially homogeneous; a dot denotes the derivative with 
respect to cosmic time $t$. Actually, in some sense our hybrid dark 
energy model is similar to the quintom 
model~\cite{r21,r22,r23,r13} or the hessence model~\cite{r19}. The
key difference is the interaction form between quintessence and 
phantom, and especially the effective potential with the feature 
of spontaneous symmetry breaking. We will see that this key 
difference makes this simple model have rich phenomenology and 
provide the possibility to alleviate the so-called second 
cosmological coincidence problem.

\begin{center}
\begin{figure}[htp]
\centering
\includegraphics[width=0.35\textwidth]{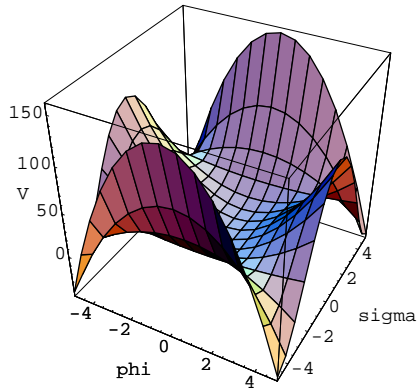}
\caption{\label{fig1} The 3D plot of the effective potential (\ref{eq5}) for 
demonstrative model parameters $g=\lambda=1$, $m=3$ and $M=0.1$.}
\end{figure}
\end{center}

Note that the quintessence {\em rolls down} its potential and 
approaches to stabilize at the {\em minimum}~\cite{r6}, while the 
phantom {\em climbs up} its potential and approaches to stabilize 
at the {\em maximum}~\cite{r10}. We show the 3D plot of the 
effective potential (\ref{eq5}) in Fig.~\ref{fig1}. To determine 
the extrema of the effective potential (\ref{eq5}), it is useful 
to work out the first and second derivatives in the $\sigma$ and 
$\phi$ directions, respectively. They are 
\bea
&&V_{,\sigma}=\left(M^2-g^2\phi^2+\lambda\sigma^2\right)\sigma\, ,\label{eq6}\\
&&V_{,\sigma\sigma}=M^2-g^2\phi^2+3\lambda\sigma^2\, ,\label{eq7}
\eea
and
\bea
&&V_{,\phi}=\left(m^2-g^2\sigma^2\right)\phi\, ,\label{eq8}\\
&&V_{,\phi\phi}=m^2-g^2\sigma^2\, ,\label{eq9} 
\eea 
where $f_{,x}\equiv\partial f/\partial x$. Following the hybrid 
inflation model~\cite{r30} and the work of Gong and 
Kim~\cite{r32}, for convenience, we only consider the case of 
$\phi \ge 0 $ and $\sigma \ge 0$, since the effective potential 
(\ref{eq5}) is symmetric. Therefore we have, 
\be{eq10}
V_{,\phi\sigma}=V_{,\sigma\phi}=-2g^2\phi\sigma  \le 0.
\ee


\begin{center}
\begin{figure}[htp]
\centering
\includegraphics[angle=90,width=0.35\textwidth]{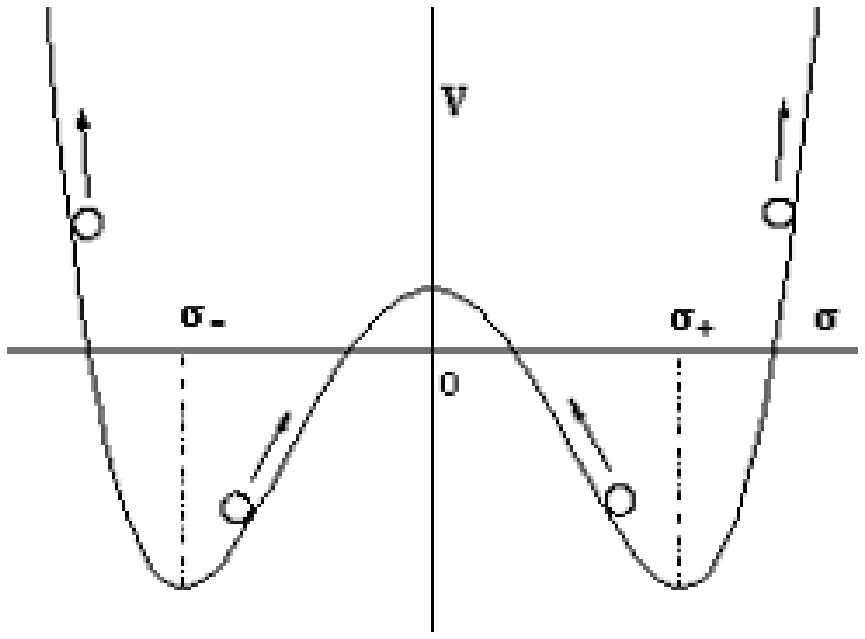}
\caption{\label{fig2} The sketch of the effective potential (\ref{eq5}) in the $\sigma$
direction for $\phi>\phi_c$.}
\end{figure}
\end{center}


For $\phi>\phi_c\equiv M/g$, from Eq.~(\ref{eq6}), we have three 
extrema for the effective potential (\ref{eq5}) in the $\sigma$ direction, i.e. 
\be{eq11}
\sigma=0~~~~~~~{\rm and}~~~~~~~
\sigma=\sigma_\pm\equiv\pm\sqrt{\frac{g^2\phi^2-M^2}{\lambda}}.
\ee
  From Eq.~(\ref{eq7}), we see that $\sigma=0$ is the only maximum while 
$\sigma=\sigma_\pm$ are minima. See Fig.~\ref{fig2}. Recall that 
the phantom {\em climbs up} its potential and approaches to 
stabilize at the {\em maximum}~\cite{r10}. To trap the field 
$\sigma$ at the maximum $\sigma=0$ in the first stage, we assume 
that the initial value of the field $\sigma$ satisfies $\sigma_-
|_{\phi=\phi_{ini}}<\sigma_{ini} <\sigma_+ |_{\phi=\phi_{ini}}$.
If the initial value of $\phi$, namely $\phi_{ini}$, is large
enough, or by appropriately choosing the parameters $g$, $\lambda$
and $M$, we can make $|\sigma_\pm|$ large enough. Therefore, the 
range $(\sigma_-,\sigma_+)$ is wide enough and needs not 
fine-tuning. Hence, in the first stage, we expect that the phantom
field $\sigma$ is trapped to $\sigma=0$, whereas the quintessence
field $\phi$ could remain large and slowly roll down its potential
for a long time. In this case, the kinetic energy term of the
field $\sigma$ vanishes, namely, $-\dot{\sigma}^2/2=0$. The
hybrid dark energy reduces to the case of a quintessence with an
effective cosmological constant. In this stage, the effective EoS
remains larger than $-1$ for a long time.


\begin{center}
\begin{figure}[htp]
\centering
\includegraphics[angle=90,width=0.35\textwidth]{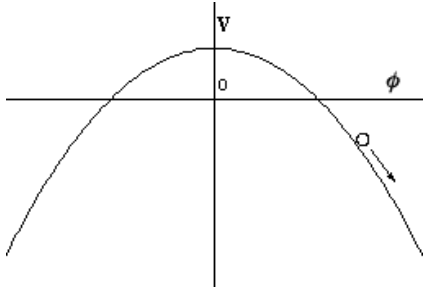}
\caption{\label{fig3} The sketch of the effective potential (\ref{eq5}) in the $\phi$
direction for $\sigma>\sigma_c$.}
\end{figure}
\end{center}


When $\phi$ becomes smaller than $\phi_c$, 
a phase transition for $\sigma$ takes place.  However, the thing 
is not so simple. From Eqs.~(\ref{eq8}) and (\ref{eq9}), $\phi=0$ 
is the minimum only when $\sigma<\sigma_c\equiv m/g$. If 
$\sigma>\sigma_c$, $\phi=0$ is maximum and the field $\phi$ 
becomes larger and larger. See Fig.~\ref{fig3}. Thus, $\phi$ 
cannot reach $\phi_c$ forever and the phase transition cannot 
happen. To realize the phase transition, we assume that 
$\sigma_c>\sigma_+ |_{\phi=\phi_{ini}}$. So, in the first stage, 
$\sigma<\sigma_c$ always holds and the field $\phi$ rolls down 
toward its minimum $\phi=0$. When $\phi<\phi_c$, from 
Eqs.~(\ref{eq6}) and (\ref{eq7}), $\sigma=0$ becomes the only  
minimum where the phantom field $\sigma$ is unstable.  As a
result, when $\phi$ reaches the critical value $\phi_c$, the phase
transition occurs and the phantom field $\sigma$ is triggered to
climb up its potential. The kinetic energy term of the phantom
field $\sigma$, namely $-\dot{\sigma}^2/2$, becomes more and more
negative as the velocity of $\sigma$ increases. On the other hand,
the quintessence field $\phi$ rolls down its potential and 
approaches to its minimum $\phi=0$ and stabilize there after some
periods of oscillation. Once $\phi$ is trapped at $\phi=0$, its
kinetic energy term $\dot{\phi}^2/2=0$. Thus, the hybrid dark
energy model reduces to the one for a pure phantom field. Therefore 
the effective EoS remains smaller than $-1$. Crossing the phantom
divide occurs between the moment of phase transition $\phi=\phi_c$
and the moment of $\phi$ being trapped at $\phi=0$ eventually.

This is not the whole story. When $\sigma$ continuously climbs up,
it will eventually arrive at the critical value $\sigma_c$
mentioned above. When $\sigma>\sigma_c$, from Eqs.~(\ref{eq8}) and
 (\ref{eq9}), $\phi=0$ becomes a maximum. The quintessence
field $\phi$ is triggered to roll down again. This is a new
feature, which does not appear in the hybrid inflation
model~\cite{r30} and the model given by Gong and Kim~\cite{r32}.
In the third stage, the quintessence field $\phi$ rolls down and
its value increases continuously. When $\phi$ reaches the critical
value $\phi_c$ once again, the motion of $\sigma$ will not change
the direction and $\sigma$ continues to climb up, since at this
moment $\sigma>\sigma_c>\sigma_+ |_{\phi=\phi_{ini}}\gg\sigma_+
|_{\phi=\phi_c}$ because of $\phi_{ini}\gg\phi_c$ [cf.
Eq.~(\ref{eq11}) and Fig.~\ref{fig2}]. In this stage, the
effective EoS is involved. Note that crossing the phantom divide 
occurs when $\dot{\phi}^2=\dot{\sigma}^2$. In this stage, the veloities 
of $\phi$ and $\sigma$ are all increasing. Therefore, whether the 
EoS remains smaller than $-1$ or changes to be larger than $-1$
depends on the profile of the effective potential (\ref{eq5}) in
the third stage.


\begin{center}
\begin{figure}[htp]
\centering
\includegraphics[width=0.49\textwidth]{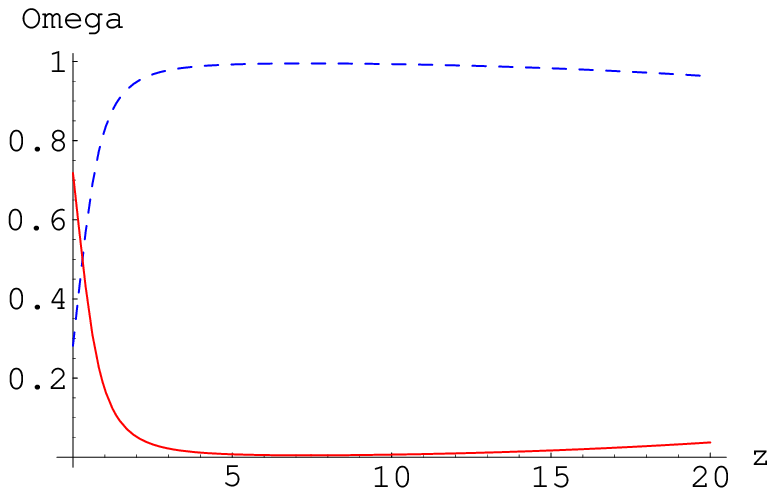}~
\includegraphics[width=0.49\textwidth]{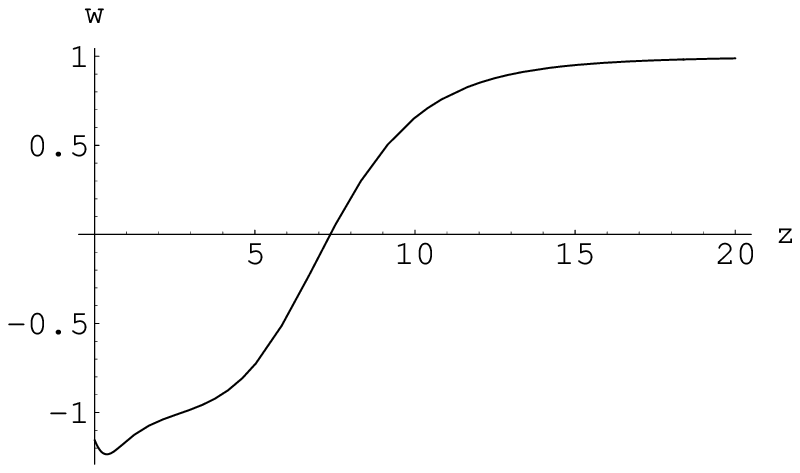}
\caption{\label{fig4} The numerical plots for the demonstrative parameters 
$\bar{g}=\bar{\lambda}=\bar{m}=\bar{M}=1$ and $\Omega_{m0}=0.3$. 
The left panel is the fractional energy densities of hybrid 
dark energy (solid line) and dust matter (dashed line) versus redshift $z$. 
The right panel is the effective EoS of hybrid dark energy versus $z$.}
\end{figure}
\end{center}

In this note we are concerned for the second stage. In this stage,
crossing the phantom divide occurs with the help of the trigger
mechanism. The effective EoS remains smaller than $-1$ in the
major part of this stage. We can choose the model parameters to
make $\sigma_c$ large enough. So, the duration of the second stage
can be long enough. Following the idea proposed in~\cite{r33}, if 
the fraction of the duration of the second stage to the lifetime
of our universe is large enough, the probability that we live in
this stage is also large enough.  On the other hand, by
appropriately choosing model parameters, we can construct a
cosmological model so that crossing the phantom divide occurs when
the dark matter energy density and dark energy density are
comparable. In this sense, the so-called second cosmological
coincidence problem can be alleviated.

To support this point, we present a numerical demonstration here. 
The equations of motion for $\phi$ and $\sigma$ are
\be{eq12}
\ddot{\phi}+3H\dot{\phi}+V_{,\phi}=0,~~~~~~~
\ddot{\sigma}+3H\dot{\sigma}-V_{,\sigma}=0,
\ee
respectively. Considering a spatially flat universe, the Friedmann equation reads 
\be{eq13}
H^2=\frac{1}{3}\left(\rho_m+\rho_X\right),
\ee
where we have used the unit $8\pi G=1$. The effective pressure and 
energy density of the hybrid dark energy are 
\be{eq14}
p_X=\frac{1}{2}\dot{\phi}^2-\frac{1}{2}\dot{\sigma}^2-V(\phi,\sigma),~~~~~~~
\rho_X=\frac{1}{2}\dot{\phi}^2-\frac{1}{2}\dot{\sigma}^2+V(\phi,\sigma),
\ee
respectively, where the effective potential $V(\phi,\sigma)$ is given by 
Eq.~(\ref{eq5}). The effective EoS of the hybrid dark energy 
$w\equiv p_X/\rho_X$. For convenience, we consider the case of minimal 
coupling between dark energy and dust matter, thus 
\be{eq15}
\rho_m=\rho_{m0}a^{-3},
\ee
where the subscript ``0'' indicates the present value of the corresponding 
quantity, $a=(1+z)^{-1}$ is the scale factor (we set $a_0=1$ throughout), and 
$z$ is the redshift. Then, we recast Eqs.~(\ref{eq12}) and (\ref{eq13}) as the 
following first-order differential equations with respect to the redshift $z$, 
\bea
&&\frac{d\phi}{dz}=-(1+z)^{-1}\bar{H}^{-1}\bar{\chi},\label{eq16}\\
&&\frac{d\sigma}{dz}=-(1+z)^{-1}\bar{H}^{-1}\bar{\eta},\label{eq17}\\
&&\frac{d\bar{\chi}}{dz}=(1+z)^{-1}\left[3\bar{\chi}+\left(\bar{m}^2-
\bar{g}^2\sigma^2\right)\phi\bar{H}^{-1}\right],\label{eq18}\\
&&\frac{d\bar{\eta}}{dz}=(1+z)^{-1}\left[3\bar{\eta}-\left(\bar{M}^2-\bar{g}^2\phi^2
+\bar{\lambda}\sigma^2\right)\sigma\bar{H}^{-1}\right],\label{eq19}
\eea
and
\be{eq20}
\bar{H}^2=\Omega_{m0}(1+z)^3+\frac{1}{3}\left[\frac{1}{2}\bar{\chi}^2-
\frac{1}{2}\bar{\eta}^2+\frac{1}{2}\bar{m}^2\phi^2-\frac{1}{2}\bar{g}^2\phi^2\sigma^2
+(4\bar{\lambda})^{-1}\left(\bar{M}^2+\bar{\lambda}\sigma^2\right)^2\right],
\ee
where $\bar{\chi}\equiv\dot{\phi}/H_0$, $\bar{\eta}\equiv\dot{\sigma}/H_0$, 
$\bar{m}\equiv m/H_0$, $\bar{M}\equiv M/H_0$, $\bar{g}\equiv g/H_0$, 
$\bar{\lambda}\equiv\lambda/H_0^2$, $\bar{H}\equiv H/H_0$ 
and $\Omega_{m0}\equiv\rho_{m0}/(3H_0^2)$. We show the numerical result 
in Fig.~\ref{fig4}. It is easy to see that the effective EoS of hybrid dark energy 
crosses the phantom divide $w=-1$ when the redshift $z$ is of order unity, 
while the fractional energy densities of dust matter and dark energy are 
comparable. \\

After all, some remarks are in order. Firstly, within the spirit of the hybrid 
inflation model~\cite{r30} and the model by Gong and 
Kim~\cite{r32}, the effective potential (\ref{eq5}) could be 
generalized. For instance, instead of the term $m^2\phi^2/2$, one 
can use the term $\mu\phi^4/4$ or other possible potentials for 
the quintessence field $\phi$. Secondly,  it is also possible in our 
model to make an alleviation for the ``first'' cosmological 
coincidence problem. Note that in the first stage the phantom 
field $\sigma$ is trapped at $\sigma=0$, the  hybrid dark energy 
model reduces to the one for a pure quintessence. As is 
well-known, it is easy to obtain a tracker solution for the 
quintessence~\cite{r7}. So, it is not strange that we are living 
in an epoch in which the energy densities of dark energy and 
matter are comparable. Thirdly, as mentioned above, in the third 
stage of our model, whether the EoS remains smaller than $-1$ or 
comes back to larger than $-1$ depends on the profile of the 
effective potential (\ref{eq5}). Thus, the avoidance of the big 
rip is also possible for suitable model parameters. Fourthly, similar 
to~\cite{r32}, one may add other scalar field into our model to unify 
the primordial inflation, present accelerated expansion and crossing 
the phantom divide. Finally, the hybrid-inflation-type models can 
arise naturally in many string/brane inspired theories (see e.g. \cite{r31} 
for a comprehensive review). And the kind of effective potentials 
(\ref{eq4}) is argued to arise from the supersymmetric theories 
in~\cite{r32}. We expect that our hybrid dark energy model in this 
note has also a solid physical foundation.


\section*{Acknowledgments}
We thank the anonymous referee for his/her quite useful comments and suggestions, 
which help us to deepen this work. H.W. is grateful to Zong-Kuan Guo, Yun-Song Piao, 
Ding-Fang Zeng, Hui Li, Xin Zhang, Li-Ming Cao, Da-Wei Pang, Yi Zhang, 
Hong-Sheng Zhang, Ding Ma, Hua Bai and Xun Su for helpful discussions. This work 
was supported in part by a grant from Chinese Academy of Sciences, and grants from 
NSFC, China (No. 10325525 and No. 90403029).



\begin{thebibliography}{99}

\bibitem{r1}
A.~G.~Riess {\it et al.} [Supernova Search Team Collaboration], Astrophys.\ J.\  {\bf 607},
665 (2004) [astro-ph/0402512];\\
R.~A.~Knop {\it et al.}, [Supernova Cosmology Project Collaboration], Astrophys.\ J.\
{\bf 598}, 102 (2003) [astro-ph/0309368];\\
A.~G.~Riess {\it et al.} [Supernova Search Team Collaboration], Astron.\ J.\  {\bf 116},
1009 (1998) [astro-ph/9805201];\\
S.~Perlmutter {\it et al.} [Supernova Cosmology Project Collaboration], Astrophys.\ J.\
{\bf 517}, 565 (1999) [astro-ph/9812133].

\bibitem{r2}
C.~L.~Bennett {\it et al.}, Astrophys.\ J.\ Suppl. {\bf 148}, 1 (2003) [astro-ph/0302207];\\
D.~N.~Spergel {\it et al.}, Astrophys.\ J.\ Suppl. {\bf 148} 175 (2003) [astro-ph/0302209].

\bibitem{r3}
M.~Tegmark {\it et al.} [SDSS Collaboration], Phys.\ Rev.\ D {\bf 69}, 103501 (2004) [astro-ph/0310723];\\
M.~Tegmark {\it et al.} [SDSS Collaboration], Astrophys.\ J.\  {\bf 606}, 702 (2004) [astro-ph/0310725];\\
U.~Seljak {\it et al.}, Phys.\ Rev.\ D {\bf 71}, 103515 (2005) [astro-ph/0407372];\\
J.~K.~Adelman-McCarthy {\it et al.} [SDSS Collaboration], astro-ph/0507711;\\
K.~Abazajian {\it et al.} [SDSS Collaboration], astro-ph/0410239; astro-ph/0403325; astro-ph/0305492.

\bibitem{r4}
S.~W.~Allen, R.~W.~Schmidt, H.~Ebeling, A.~C.~Fabian and L.~van Speybroeck, Mon.\ Not.\ Roy.\ Astron.\ Soc.\  {\bf 353}, 457 (2004)
[astro-ph/0405340].

\bibitem{r5}
P.~J.~E.~Peebles and B.~Ratra, Rev.\ Mod.\ Phys.\  {\bf 75}, 559 (2003) [astro-ph/0207347];\\
T.~Padmanabhan, Phys.\ Rept.\  {\bf 380}, 235 (2003) [hep-th/0212290];\\
S.~M.~Carroll, astro-ph/0310342;\\
R.~Bean, S.~Carroll and M.~Trodden, astro-ph/0510059;\\
V.~Sahni and A.~A.~Starobinsky, Int.\ J.\ Mod.\ Phys.\ D {\bf 9}, 373 (2000) [astro-ph/9904398];\\
S.~M.~Carroll, Living Rev.\ Rel.\  {\bf 4}, 1 (2001) [astro-ph/0004075];\\
T.~Padmanabhan, Curr.\ Sci.\  {\bf 88}, 1057 (2005) [astro-ph/0411044];\\
S.~Weinberg, Rev.\ Mod.\ Phys.\  {\bf 61}, 1 (1989);\\
S.~Nobbenhuis, gr-qc/0411093.

\bibitem{r6}
R.~R.~Caldwell, R.~Dave and P.~J.~Steinhardt, Phys.\ Rev.\ Lett.\  {\bf 80}, 1582 (1998) [astro-ph/9708069];\\
C.~Wetterich, Nucl.\ Phys.\ B {\bf 302}, 668 (1988);\\
P.~J.~E.~Peebles and B.~Ratra, Astrophys.\ J.\  {\bf 325}, L17 (1988).

\bibitem{r7}
P.~J.~Steinhardt, L.~M.~Wang and I.~Zlatev, Phys.\ Rev.\ D {\bf 59}, 123504 (1999) [astro-ph/9812313];\\
I.~Zlatev and P.~J.~Steinhardt, Phys.\ Lett.\ B {\bf 459}, 570 (1999) [astro-ph/9906481].

\bibitem{r8}
R.~R.~Caldwell, Phys.\ Lett.\ B {\bf 545}, 23 (2002) [astro-ph/9908168];\\
R.~R.~Caldwell, M.~Kamionkowski and N.~N.~Weinberg, Phys.\ Rev.\ Lett.\  {\bf 91}, 071301 (2003) [astro-ph/0302506].

\bibitem{r9}
S.~M.~Carroll, M.~Hoffman and M.~Trodden, Phys.\ Rev.\ D {\bf 68}, 023509 (2003) [astro-ph/0301273];\\
J.~M.~Cline, S.~Jeon and G.~D.~Moore, Phys.\ Rev.\ D {\bf 70}, 043543 (2004) [hep-ph/0311312].

\bibitem{r10}
Y.~S.~Piao and E.~Zhou, Phys.\ Rev.\ D {\bf 68}, 083515 (2003) [hep-th/0308080];\\
Y.~S.~Piao and Y.~Z.~Zhang, Phys.\ Rev.\ D {\bf 70}, 063513 (2004) [astro-ph/0401231];\\
Z.~K.~Guo, Y.~S.~Piao and Y.~Z.~Zhang, Phys.\ Lett.\ B {\bf 594}, 247 (2004) [astro-ph/0404225].

\bibitem{r11}
D.~Huterer and A.~Cooray, Phys.\ Rev.\ D {\bf 71}, 023506 (2005) [astro-ph/0404062];\\
U.~Alam, V.~Sahni and A.~A.~Starobinsky, JCAP {\bf 0406}, 008 (2004) [astro-ph/0403687];\\
Y.~Wang and M.~Tegmark, Phys.\ Rev.\ D {\bf 71}, 103513 (2005) [astro-ph/0501351];\\
R.~Lazkoz, S.~Nesseris and L.~Perivolaropoulos, astro-ph/0503230.

\bibitem{r12}
A.~Upadhye, M.~Ishak and P.~J.~Steinhardt, Phys.\ Rev.\ D {\bf 72}, 063501 (2005) [astro-ph/0411803].

\bibitem{r13}
B.~Feng, X.~L.~Wang and X.~M.~Zhang, Phys.\ Lett.\ B {\bf 607}, 35 (2005) [astro-ph/0404224].

\bibitem{r14}
W.~Hu, Phys.\ Rev.\ D {\bf 71}, 047301 (2005) [astro-ph/0410680].

\bibitem{r15}
R.~R.~Caldwell and M.~Doran, Phys.\ Rev.\ D {\bf 72}, 043527 (2005) [astro-ph/0501104];\\
E.~Elizalde, S.~Nojiri and S.~D.~Odintsov, Phys.\ Rev.\ D {\bf 70}, 043539 (2004) [hep-th/0405034];\\
B.~McInnes, Nucl.\ Phys.\ B {\bf 718}, 55 (2005) [hep-th/0502209];\\
H.~Stefancic, Phys.\ Rev.\ D {\bf 71}, 124036 (2005) [astro-ph/0504518];\\
S.~Nojiri and S.~D.~Odintsov, Phys.\ Rev.\ D {\bf 72}, 023003 (2005) [hep-th/0505215].

\bibitem{r16}
S.~Nojiri, S.~D.~Odintsov and S.~Tsujikawa, Phys.\ Rev.\ D {\bf 71}, 063004 (2005) [hep-th/0501025].

\bibitem{r17}
Y.~H.~Wei and Y.~Z.~Zhang, Grav.\ Cosmol.\  {\bf 9}, 307 (2003) [astro-ph/0402515];\\
Y.~H.~Wei and Y.~Tian, Class.\ Quant.\ Grav.\  {\bf 21}, 5347 (2004) [gr-qc/0405038];\\
Y.~H.~Wei, Mod.\ Phys.\ Lett.\ A {\bf 20}, 1147 (2005) [gr-qc/0410050];\\
Y.~H.~Wei, gr-qc/0502077.

\bibitem{r18}
R.~G.~Cai, H.~S.~Zhang and A.~Wang, Commun.\ Theor.\ Phys.\  {\bf 44}, 948 (2005) [hep-th/0505186];\\
V.~Sahni, astro-ph/0502032;\\
V.~Sahni and Y.~Shtanov, JCAP {\bf 0311}, 014 (2003) [astro-ph/0202346];\\
I.~Y.~Aref'eva, A.~S.~Koshelev and S.~Y.~Vernov, Phys.\ Rev.\ D {\bf 72}, 064017 (2005) [astro-ph/0507067].

\bibitem{r19}
H.~Wei, R.~G.~Cai and D.~F.~Zeng, Class.\ Quant.\ Grav.\  {\bf 22}, 3189 (2005) [hep-th/0501160];\\
H.~Wei and R.~G.~Cai, Phys.\ Rev.\ D {\bf 72}, 123507 (2005) [astro-ph/0509328].

\bibitem{r20}
A.~Vikman, Phys.\ Rev.\ D {\bf 71}, 023515 (2005) [astro-ph/0407107].

\bibitem{r21}
Z.~K.~Guo, Y.~S.~Piao, X.~M.~Zhang and Y.~Z.~Zhang, Phys.\ Lett.\ B {\bf 608}, 177 (2005) [astro-ph/0410654].

\bibitem{r22}
X.~F.~Zhang, H.~Li, Y.~S.~Piao and X.~M.~Zhang, astro-ph/0501652.

\bibitem{r23}
G.~B.~Zhao, J.~Q.~Xia, M.~Z.~Li, B.~Feng and X.~M.~Zhang, astro-ph/0507482.

\bibitem{r24}
M.~Z.~Li, B.~Feng and X.~M.~Zhang, hep-ph/0503268;\\
A.~Anisimov, E.~Babichev and A.~Vikman, JCAP {\bf 0506}, 006 (2005) [astro-ph/0504560].

\bibitem{r25}
X.~Zhang and F.~Q.~Wu, Phys.\ Rev.\ D {\bf 72}, 043524 (2005) [astro-ph/0506310];\\
X.~Zhang, Int.\ J.\ Mod.\ Phys.\ D {\bf 14}, 1597 (2005) [astro-ph/0504586];\\
B.~Wang, Y.~G.~Gong and E.~Abdalla, Phys.\ Lett.\ B {\bf 624}, 141 (2005) [hep-th/0506069].

\bibitem{r26}
L.~Perivolaropoulos, astro-ph/0504582;\\
M.~X.~Luo and Q.~P.~Su, astro-ph/0506093;\\
F.~C.~Carvalho and A.~Saa, Phys.\ Rev.\ D {\bf 70}, 087302 (2004) [astro-ph/0408013];\\
V.~Faraoni, Phys.\ Rev.\ D {\bf 70}, 044037 (2004) [gr-qc/0407021];\\
V.~Faraoni, E.~Gunzig and P.~Nardone, Fund.\ Cosmic Phys.\  {\bf 20}, 121 (1999) [gr-qc/9811047].

\bibitem{r27}
C.~G.~Huang and H.~Y.~Guo, astro-ph/0508171.

\bibitem{r28}
H.~S.~Zhang and Z.~H.~Zhu, astro-ph/0509895.

\bibitem{r29}
R.~G.~Cai, Y.~G.~Gong and B.~Wang, hep-th/0511301.

\bibitem{r30}
A.~D.~Linde, Phys.\ Rev.\ D {\bf 49}, 748 (1994) [astro-ph/9307002];\\
A.~D.~Linde, Phys.\ Lett.\ B {\bf 259}, 38 (1991).

\bibitem{r31}
F.~Quevedo, Class.\ Quant.\ Grav.\  {\bf 19}, 5721 (2002) [hep-th/0210292].

\bibitem{r32}
J.~O.~Gong and S.~Kim, astro-ph/0510207.

\bibitem{r33}
R.~J.~Scherrer, Phys.\ Rev.\ D {\bf 71}, 063519 (2005) [astro-ph/0410508];\\
R.~G.~Cai and A.~Wang, JCAP {\bf 0503}, 002 (2005) [hep-th/0411025];\\
B.~McInnes, astro-ph/0210321.

\bibitem{r34}
S.~Das, P.~S.~Corasaniti and J.~Khoury, astro-ph/0510628.

\bibitem{r35}
S.~Nojiri and S.~D.~Odintsov, hep-th/0506212;\\
A.~A.~Andrianov, F.~Cannata and A.~Y.~Kamenshchik, gr-qc/0512038;\\
A.~A.~Andrianov, F.~Cannata and A.~Y.~Kamenshchik, Phys.\ Rev.\ D {\bf 72}, 043531 (2005) [gr-qc/0505087].

\bibitem{r36}
M.~Axenides and K.~Dimopoulos, JCAP {\bf 0407}, 010 (2004) [hep-ph/0401238].

\bibitem{r37}
P.~F.~Gonzalez-Diaz, Int.\ J.\ Mod.\ Phys.\ D {\bf 14}, 1313 (2005);\\
P.~F.~Gonzalez-Diaz, Phys.\ Rev.\ D {\bf 69}, 063522 (2004) [hep-th/0401082];\\
P.~F.~Gonzalez-Diaz, Phys.\ Lett.\ B {\bf 586}, 1 (2004) [astro-ph/0312579].

\bibitem{r38}
S.~Capozziello, S.~Nojiri and S.~D.~Odintsov, hep-th/0507182;\\
B.~M.~N.~Carter and I.~P.~Neupane, hep-th/0510109;\\
B.~M.~N.~Carter and I.~P.~Neupane, hep-th/0512262.

\end{thebibliography}
\end{document}